\documentclass[fleqn,usenatbib]{mnras}

\usepackage{newtxtext,newtxmath}

\usepackage[T1]{fontenc}

\DeclareRobustCommand{\VAN}[3]{#2}
\let\VANthebibliography\thebibliography
\def\thebibliography{\DeclareRobustCommand{\VAN}[3]{##3}\VANthebibliography}

\usepackage{graphicx}	
\usepackage{amsmath}	
\usepackage{multirow}

\usepackage{orcidlink}

\title[0.3\% calibration of CB luminosity]{A 0.3\% calibration of the W UMa-type contact binary luminosity based on Gaia DR3 }

\author[Jing Li et al.]{
Jing Li,$^{1,2}$
Xiaodian Chen\orcidlink{0000-0001-7084-0484},$^{2,1,3,4}$\thanks{E-mail: chenxiaodian@nao.cas.cn} 
Shu Wang,$^{2,1}$
Kun Wang,$^{1}$
Kai Li,$^{5}$
Licai Deng,$^{2,1,3}$
\\
$^{1}$School of Physics and Astronomy, China West Normal University, Nanchong 637009, China\\
$^{2}$CAS Key Laboratory of Optical Astronomy, National Astronomical Observatories, Chinese Academy of Sciences, Beijing 100101,  China\\
$^{3}$School of Astronomy and Space Science, University of the Chinese Academy of Sciences, Beijing, 100049, China\\
$^{4}$Institute for Frontiers in Astronomy and Astrophysics, Beijing Normal University,  Beijing 102206, China\\
$^{5}$Shandong Key Laboratory of Optical Astronomy and Solar-Terrestrial Environment, School of Space Science and Physics, \\
Institute of Space Sciences, Shandong University, Weihai, Shandong, 264209, China}

\date{Accepted XXX. Received YYY; in original form ZZZ}

\pubyear{2025}

\begin{document}
\label{firstpage}
\pagerange{\pageref{firstpage}--\pageref{lastpage}}
\maketitle

\begin{abstract}

W Ursa Majoris (W UMa)-type contact binary systems (CBs) with period--luminosity (PL) relations are valuable distance indicators. The PL relations of CBs are affected by metallicity. Here, we establish PL relations and period--luminosity--metallicity (PLZ) relations in nine bands from optical to mid-infrared ($BP$, $G$, $RP$, $J$, $H$, $K_S$, $W1$, $W2$, $W3$) and in five Wesenheit bands based on Gaia DR3 parallaxes. The dispersion of PLZ relations gradually decreases from the optical to mid-infrared bands, with the minimum dispersion of 0.138 mag. We fit the best PL relations for three bands ($W1$, $W_{G,BP,RP}$, $W_{W1,BP,RP}$) under different parallax uncertainty criteria and determine a parallax (after correction) zero point of $zp_\varpi=24\pm4$ $\mu$as. After fixing the parallax zero point, we find that the total zero errors of the PL and PLZ relation are smallest when the parallax uncertainty is less than 2\%, resulting in a calibrated CB luminosity with an accuracy of 0.33\%, which is more accurate than the classical Cepheids. Furthermore, by examining the absolute magnitude residuals in different metallicity intervals, we find that the use of a linear metallicity effect is appropriate for CBs with different metallicities. These results indicate that CBs are excellent standard candles.

\end{abstract}


\section{Introduction} \label{sec:intro}
Eclipsing binaries are relatively easy to detect due to their characteristic light curves, most of which are contact binary systems (CBs). \citet{kuiper1941interpretation} identified a discrepancy—later known as the ``Kuiper Paradox''—between the mass--radius relation for zero-age stars in CBs and the size of the Roche lobe relative to stellar mass. To address this, \citet{lucy1968structure} proposed a model for zero-age CBs with a common convective envelope, which successfully explained the observed light curves of W UMa-type CBs. \cite{eggen1978contact} introduced an evolutionary framework for CBs, where CBs consist of two closely orbiting stars, each overflowing its Roche lobe, forming a shallow common envelope \citep{smith1984theory}. 
Building on this, \cite{hilditch1988evolutionary} revealed distinct evolutionary pathways for shallow-contact W-type and deep-contact A-type CBs and presented the first empirical evidence of systematic shifts in effective temperature and luminosity between components due to interstellar brightness transfer, consistent with theoretical expectations. Based on empirical data, \citet{hilditch1989evolution} suggested that short-period binaries can evolve into contact systems through either direct contact or case A mass transfer. However, the field continues to face challenges, including a scarcity of high-quality spectroscopic data---especially for low-mass systems with EB-type light curves---and the requirement for long-term monitoring to determine mass transfer rates.
To address CB evolution, \citet{gazeas2008angular} analyzed over 100 cool CBs and proposed a model in which angular momentum loss through stellar winds and mass transfer from evolved A-type components drive systems toward W- or A-type CB configurations. These processes may eventually result in a merger into a rapidly rotating single star.
Furthermore, based on two- and three-dimensional correlation analyses, \citet{gazeas2009physical} demonstrated that physical parameters such as mass, radius, and luminosity are tightly linked to orbital period and mass ratio in three-dimensional space, which can be used both to validate parameter solutions and to rapidly estimate values for sky survey targets. A period dividing line at $\log P = -0.25$ separates W UMa-type CBs ($\log P < -0.25$) from early-type CBs ($\log P > -0.25$), with the former exhibiting a steeper period--luminosity (PL) relation \citep{chen2016contact}. Given the large population of W UMa-type CBs, they have been proposed as potential distance indicators \citep{chen2016contact}.

\cite{rucinski1974luminosities} and \cite{mochnacki1981contact} were the first to study the PL relations of W UMa-type CBs. Over the past decades, multiband PL relations have been investigated for both W UMa-type and early-type CBs. 
For example, \cite{muraveva2014eclipsing} investigated the PL relations in the $K_S$ band. Based on 66 CBs with Hipparcos parallaxes, \cite{chen2016contact} derived PL relations for W UMa-type CBs in the near-infrared $J$, $H$ and $K_{S}$ bands. Additionally, for the first time, they derived PL relations for early-type CBs. They suggested that CBs can be used as distance tracers. It was also found that in the near-infrared band, CBs fainter by 0.2 mag for every 1.0 dex increase in metallicity. Later, with the Tycho-Gaia Astrometric Solution (TGAS) data, a lot of work focused on PL and PLC relations of CBs. \cite{mateo2017absolute} analyzed the effects of metallicity on absolute magnitudes of W UMa-type CBs and calibrated the near-infrared band PL and PLC relations. Based on 183 W UMa-type CBs with accurate Tycho-Gaia parallaxes, \cite{chen2018optical} explored PL relations in a total of 12 bands from optical to mid-infrared, achieving a minimum scatter of 0.16 mag and a distance precision of 7\%. \cite{jayasinghe2020asas} derived the optical to mid-infrared PL relations for both the W UMa-type and the early-type CBs. \cite{ngeow2021zwicky} derived the PL and PLC relations for W UMa-type CBs in the g- and r- bands. \cite{song2024period} proposed fine relations of period--luminosity--metallicity-color (PLZC)  from ultraviolet to mid-infrared bands, with accuracies of 6\% and 8\% in the infrared and optical bands, respectively. There is still a lack of detailed studies of PLZ relations for CB stars in previous work. In particular, how large the zero point error of the relation is has not been investigated, which is particularly important for distance measurements using the CBs. In contrast, for Cepheids and RR Lyrae stars, studies of PLZ relations are more common and have yielded remarkable results, with reference in \citep{marconi2015new,breuval2022improved,neeley2019standard,riess2021cosmic}. Therefore, for CB stars, which can also be used as standard candles, we study their PLZ relation zero points in detail in this work. Furthermore, in order to obtain more precise PLZ relations, we chose to study CBs using a parallax space fitting method based on the Gaia parallax.

\cite{lindegren2018gaia} defines the term ‘parallax zero point’ using quasar data from Gaia Data Release 2 \citep{brown2018gaia}. However, it was found to be not a single value due to variations in source colors or magnitudes, hence referred to as "parallax offset". \cite{riess2018milky} discovered that Gaia parallax offset can be measured based on Cepheids, although with low precision. \cite{riess2021cosmic} used the Gaia EDR3 parallax \citep{brown2021gaia} to calibrate the parallax offset and the PL relation simultaneously, resulting in a distance accuracy of 1\% for the Cepheid PL relation. Furthermore, the study revealed the dependency of the Cepheids PL relations and its zero point on metallicity. The study of RR Lyrae stars by \cite{neeley2019standard} has shown that metallicity plays a crucial role in the zero point, with metal-rich stars fainter than metal-poor ones in the same band, and that metallicity has a greater effect on the optical band than on the near-infrared band. 

By using a more accurate fitting method, a larger LAMOST metallicity dataset, and accurate Gaia parallaxes, we can perform better studies on CB PLZ relations. We present the CB data used in this study in Section \ref{sec:data} and discuss the methods and results in Section \ref{sec:style}. We present the discussion and conclusions in Sections \ref{sec:floats} and \ref{sec:5}.

\section{Data} \label{sec:data}

\subsection{Sample} 
The eclipsing binary catalog of Gaia Data Release 3 (DR3) contains more than 2 million candidates \citep{2023A&A...674A..16M}. To select CBs, we chose the sample labelled as `2G-B' and `2G-C'. To further purify the sample, we required that the difference in depth between the primary and secondary eclipses be less than 0.1 mag, which gave us 680,000 CBs. We then crossed the samples with the Large Sky Area Multi-Object Fiber Spectroscopic Telescope (LAMOST)  data \citep{luo2015first} and obtained 6316 CBs with metallicity [Fe/H].
We cross-matched the sample with Gaia DR3 to obtain the parallax and photometry in the $BP$, $G$, and $RP$ bands. Then, we continued to cross-match the obtained data with the Two Micron
 All Sky Survey \citep[2MASS,][]{skrutskie2006two}, to obtain photometry in the $J$, $H$, and $K_{S}$ bands, and with Wide-field Infrared Survey Explorer \citep[WISE,][]{wright2010wide} to obtain photometry in the $W1$, $W2$, and $W3$ bands. With the code\footnote{\url{https://pypi.org/project/gaiadr3-zeropoint/}}, we determined the parallax offset and then calculated the corrected parallax for our sample. The color excess $E(B-V)$ was obtained by cross-matching with the 3D dust reddening map\footnote{\href{http://argonaut.skymaps.info/} {http://argonaut.skymaps.info/}} \citep{green20193d}. Through these processes, we have obtained a total of 5972 samples.

For this sample, we further impose quality restrictions:

1. We require RUWE $<$ 1.4, an astrometric quality parameter used to help identify spurious parallaxes through Gaia. 

2. Corrected-parallax $\varpi > 0$ and parallax uncertainty ${\rm (e\_plx) / plx} < 2\%$.

3. For CB stars, we selected stars with photometric errors $<$ 0.03, 0.03, 0.55, 0.07, 0.04, and 0.07 mag in the $W1$, $W2$, $W3$, $BP$, $G$ and $RP$ bands \citep{song2024period}. In order to have reliable photometric magnitudes, we have therefore taken the limiting magnitudes of 15.0, 15.0, and 11.0 for the $W1$, $W2$, $W3$ bands as the bright direction limits \citep{chen2018optical}.

4. The sample was restricted to 3$\sigma$ based on the derived PL relations.

Eventually, in the $BP$ band we obtained 1215 CBs that satisfy the above criteria. Since W UMa-type and early-type CBs follow different PL relations, we divide the sample at $\log P = -0.25$, classifying the CBs into 1051 W UMa-type CBs and 164 early-type CBs. The final number of CBs in each band is listed in Table \ref{tab1}. 

\begin{table*}
    \scriptsize
    \centering
    \caption{PL and PLZ relations for W UMa-type and early-type CBs determined based on a Gaia parallax zero point of \( zp_\varpi = 24\pm 4\) $\mu$as.}
    \begin{tabular}{cccccccc}
    \hline 
     & Filters & $a_1$ & $a_2$ & $a_3$ & Total zp error \( zp_{\rm tot} \) (\%) & $\sigma$ & n\\
    \hline
    \multicolumn{8}{c}{\textbf{\emph{$M_\lambda = a_2 + a_1(\log P + 0.45) \quad$}}}\\
    & $ BP $ & $-10.168\pm0.185$ & $4.365\pm0.015$ &  & $0.73$ & $0.444$ & $1056$ \\
    & $ G $ & $-9.478\pm0.154$ & $4.046\pm0.013$ &  & $ 0.63 $ & $0.372$ & $1054$ \\
    & $ RP $ & $-8.627\pm0.126$ & $3.534\pm0.010$ &  & $ 0.54 $ & $0.308$ & $1053$ \\
    & $ J $ & $-7.430\pm0.090$ & $2.943\pm0.007$ &  & $ 0.43 $ & $0.219$ & $1040$ \\
    & $ H $ & $-6.411\pm0.070$ & $2.630\pm0.006$ &  & $ 0.37 $ & $0.172$ & $1031$ \\
    & $ K_S $ & $-6.236\pm0.066$ & $2.547\pm0.006$ &  & $ 0.36 $ & $0.162$ & $1031$ \\
    & $ W1 $ & $-6.183\pm0.060$ & $2.470\pm0.005$ &  & $ 0.34 $ & $0.145$ & $1023$ \\
    & $ W2 $ & $-6.231\pm0.061$ & $2.493\pm0.005$ &  & $ 0.34 $ & $0.149$ & $1027$ \\
    & $ W3 $ & $-6.091\pm0.174$ & $2.394\pm0.016$ &  & $ 0.75 $ & $0.209$ & $228$ \\
    & $ W_{H,K_S} $ & $-5.950\pm0.065$ & $2.425\pm0.005$ &  & $ 0.35 $ & $0.160$ & $1033$ \\
    & $ W_{J,K_S} $ & $-5.670\pm0.059$ & $2.359\pm0.005$ &  & $ 0.35 $ & $0.147$ & $1027$ \\
    & $ W_{J,H} $ & $-5.251\pm0.057$ & $2.256\pm0.005$ &  & $ 0.34 $ & $0.142$ & $1023$ \\
    & $ W_{W1,BP,RP} $ & $-6.007\pm0.057$ & $2.387\pm0.005$ &  & $ 0.33 $ & $0.139$ & $1028$ \\
    $\log P <-0.25$ & $ W_{G,BP,RP} $ & $-6.325\pm0.056$ & $2.401\pm0.005$ &  & $ 0.33 $ & $0.137$ & $1029$ \\
    \multicolumn{8}{c}{\textbf{\emph{$M_\lambda = a_2 + a_1(\log P + 0.45) + a_3[\mathrm{Fe/H}] \quad$}}} \\
   (W UMa-type) & $ BP $ & $-10.622\pm0.173$ & $4.482\pm0.015$ & $0.774\pm0.054$ & $0.73$ & $0.404$ & $1056$ \\
    & $ G $ & $-9.846\pm0.144$ & $4.134\pm0.013$ & $0.627\pm0.045$ & $ 0.63 $ & $0.340$ & $1054$ \\
    & $ RP $ & $-8.916\pm0.119$ & $3.600\pm0.010$ & $0.488\pm0.037$ & $ 0.54 $ & $0.282$ & $1053$ \\
    & $ J $ & $-7.607\pm0.087$ & $2.982\pm0.008$ & $0.307\pm0.027$ & $ 0.43 $ & $0.205$ & $1040$ \\
    & $ H $ & $-6.522\pm0.069$ & $2.653\pm0.006$ & $0.193\pm0.022$ & $ 0.37 $ & $0.165$ & $1031$ \\
    & $ K_S $ & $-6.323\pm0.065$ & $2.565\pm0.006$ & $0.154\pm0.021$ & $ 0.36 $ & $0.156$ & $1031$ \\
    & $ W1 $ & $-6.276\pm0.059$ & $2.490\pm0.005$ & $0.166\pm0.018$ & $ 0.34 $ & $0.138$ & $1023$ \\
    & $ W2 $ & $-6.339\pm0.059$ & $2.516\pm0.005$ & $0.192\pm0.019$ & $ 0.34 $ & $0.140$ & $1027$ \\
    & $ W3 $ & $-6.202\pm0.170$ & $2.420\pm0.016$ & $0.249\pm0.055$ & $ 0.75 $ & $0.202$ & $228$ \\
    & $ W_{H,K_S} $ & $-6.014\pm0.065$ & $2.438\pm0.006$ & $0.116\pm0.020$ & $ 0.36 $ & $0.157$ & $1033$ \\
    & $ W_{J,K_S} $ & $-5.727\pm0.060$ & $2.370\pm0.005$ & $0.099\pm0.019$ & $ 0.35 $ & $0.145$ & $1027$ \\
    & $ W_{J,H} $ & $-5.294\pm0.057$ & $2.265\pm0.005$ & $0.076\pm0.018$ & $ 0.34 $ & $0.140$ & $1023$ \\
    & $ W_{W1,BP,RP} $ & $-6.087\pm0.056$ & $2.403\pm0.005$ & $0.140\pm0.018$ & $ 0.33 $ & $0.134$ & $1028$ \\
    & $ W_{G,BP,RP} $ & $-6.393\pm0.056$ & $2.416\pm0.005$ & $0.121\pm0.018$ & $ 0.33 $ & $0.133$ & $1029$ \\
    \hline
    \multicolumn{8}{c}{\textbf{\emph{$M_\lambda = a_2 + a_1(\log P + 0.10) \quad$}}} \\
    & $ BP $ & $-2.492\pm0.464$ & $2.321\pm0.043$ &  & $ 2.00 $ & $0.457$ & $178$ \\
    & $ G $ & $-2.617\pm0.406$ & $2.098\pm0.037$ &  & $ 1.76 $ & $0.404$ & $179$ \\
    & $ RP $ & $-2.766\pm0.346$ & $1.703\pm0.032$ &  & $ 1.50 $ & $0.343$ & $179$ \\
    & $ J $ & $-3.050\pm0.299$ & $1.237\pm0.027$ &  & $ 1.31 $ & $0.301$ & $185$ \\
    & $ H $ & $-3.307\pm0.302$ & $1.042\pm0.028$ &  & $ 1.32 $ & $0.310$ & $189$ \\
    & $ K_S $ & $-3.355\pm0.295$ & $0.970\pm0.027$ &  & $ 1.29 $ & $0.304$ & $190$ \\
    & $ W1 $ & $-3.527\pm0.286$ & $0.884\pm0.026$ &  & $ 1.25 $ & $0.289$ & $187$ \\
    & $ W2 $ & $-3.532\pm0.282$ & $0.902\pm0.026$ &  & $ 1.24 $ & $0.286$ & $187$ \\
    & $ W3 $ & $-4.174\pm0.351$ & $0.714\pm0.031$ &  & $ 1.46 $ & $0.252$ & $88$ \\
    & $ W_{H,K_S} $ & $-3.505\pm0.283$ & $0.852\pm0.026$ &  & $ 1.25 $ & $0.289$ & $187$ \\
    & $ W_{J,K_S} $ & $-3.665\pm0.272$ & $0.816\pm0.025$ &  & $ 1.20 $ & $0.274$ & $186$ \\
    & $ W_{J,H} $ & $-3.636\pm0.273$ & $0.781\pm0.025$ &  & $ 1.20 $ & $0.270$ & $186$ \\
    & $ W_{W1,BP,RP} $ & $-3.683\pm0.276$ & $0.805\pm0.025$ &  & $ 1.21 $ & $0.279$ & $186$ \\
    $\log P > -0.25$ & $ W_{G,BP,RP} $ & $-3.446\pm0.318$ & $0.857\pm0.029$ &  & $ 1.37 $ & $0.332$ & $191$ \\
    \multicolumn{8}{c}{\textbf{\emph{$M_\lambda = a_2 + a_1(\log P + 0.10) + a_3[\mathrm{Fe/H}] \quad$}}} \\
    (early-type) & $ BP $ & $-2.488\pm0.471$ & $2.320\pm0.043$ & $-0.010\pm0.155$ & $2.03$ & $0.457$ & $178$ \\
    & $ G $ & $-2.573\pm0.411$ & $2.094\pm0.038$ & $-0.100\pm0.135$ & $ 1.78 $ & $0.404$ & $179$ \\
    & $ RP $ & $-2.681\pm0.347$ & $1.696\pm0.032$ & $-0.186\pm0.114$ & $ 1.51 $ & $0.340$ & $179$ \\
    & $ J $ & $-2.961\pm0.301$ & $1.232\pm0.027$ & $-0.172\pm0.098$ & $ 1.31 $ & $0.298$ & $185$ \\
    & $ H $ & $-3.179\pm0.303$ & $1.034\pm0.028$ & $-0.235\pm0.097$ & $ 1.32 $ & $0.306$ & $189$ \\
    & $ K_S $ & $-3.227\pm0.296$ & $0.962\pm0.027$ & $-0.229\pm0.095$ & $ 1.29 $ & $0.300$ & $190$ \\
    & $ W1 $ & $-3.401\pm0.286$ & $0.877\pm0.026$ & $-0.235\pm0.091$ & $ 1.24 $ & $0.284$ & $187$ \\
    & $ W2 $ & $-3.412\pm0.283$ & $0.894\pm0.026$ & $-0.225\pm0.090$ & $ 1.23 $ & $0.282$ & $187$ \\
    & $ W3 $ & $-4.221\pm0.358$ & $0.719\pm0.032$ & $0.095\pm0.116$ & $ 1.49 $ & $0.252$ & $88$ \\
    & $ W_{H,K_S} $ & $-3.389\pm0.286$ & $0.847\pm0.026$ & $-0.192\pm0.090$ & $ 1.24 $ & $0.286$ & $187$ \\
    & $ W_{J,K_S} $ & $-3.543\pm0.273$ & $0.809\pm0.025$ & $-0.207\pm0.086$ & $ 1.19 $ & $0.270$ & $186$ \\
    & $ W_{J,H} $ & $-3.505\pm0.274$ & $0.775\pm0.025$ & $-0.220\pm0.086$ & $ 1.20 $ & $0.265$ & $186$ \\
    & $ W_{W1,BP,RP} $ & $-3.556\pm0.276$ & $0.799\pm0.025$ & $-0.226\pm0.087$ & $ 1.21 $ & $0.274$ & $186$ \\
    & $ W_{G,BP,RP} $ & $-3.229\pm0.311$ & $0.847\pm0.028$ & $-0.393\pm0.100$ & $ 1.34 $ & $0.321$ & $191$ \\
    \hline
    \end{tabular}
    \label{tab1}
\end{table*}

\subsection{Method}
\label{sec:2.2}
In this section, we derive the PL and PLZ relations for the nine bands and the Period--Wesenheit (PW) relations. The PW relation is based on the Wesenheit magnitude. Wesenheit magnitudes \citep{madore1982period}, usually denoted by $W$, are used to reduce the interstellar extinction effect by combining two or three bands. It is commonly used as $W = V - R(V - I)$ where \(V\) and \(I\) represent the apparent magnitudes in the two bands and \(R\) is the extinction coefficient. We consider a total of five Wesenheit magnitudes, including $W_{H,K_S}$, $W_{J,K_S}$, $W_{J,H}$, $W_{W1,BP,RP}$, $W_{G,BP,RP}$. Based on the calculation for Wesenheit bands in \cite{chen2020zwicky,wang2019optical}, we derive the $R$ for the five Wesenheit bands as 1.465, 0.470, 1.164, 1.890 and 0.094. For obtaining the PL relations, the slope and intercept of the PL relations calculated in "parallax space" are more accurate due to the fact that the dominant error is normally distributed in that space. In this work, the estimation of all parameters is done in parallax space, while the presentation of the figures is done in magnitude space. Based on the PL relations for Cepheids presented in \cite{riess2021cosmic}, we obtain the expression for the PL relations of $i$ CBs as
 \begin{equation}
\mu_{0,i}=m_{W,i}-(M_{W,\log\,P_{0}}+b_W\, (\log\,P_{i}+\log\,P_{0})+Z_W \ {\rm [Fe/H]}_{i}). \quad  
\end{equation}

The mean period $\log\,P_0$ refers to the average of the $\log P$ values of the sample used for fitting the PL relation. It has statistical significance in that, when $\log P = \log P_0$, the fitted intercept directly represents the zero point of the PL relation. If $\log P$ is not centered by subtracting $\log P_0$, the intercept would be affected by the period coefficient, making its interpretation less straightforward. For W UMa-type CBs, we adopt a mean period of $\log\,P_0 = -0.45$; for early-type CBs, $\log\,P_0 = -0.10$.

$M_{W,\log\,P_0}$ is the absolute magnitude at $\log P = \log P_0$ (also called zero point of PL relation), and $\mu_{0,i}$ is the distance modulus of the $i$-th CB. $b_W$ and $Z_W$ are the coefficients for period $\log P$ and metallicity $\rm [Fe/H]$, respectively. The corresponding photometric parallax, in units of milliarcseconds (mas), is given by
\begin{equation}
\pi_{\mathrm{phot},i} = 10^{-0.2(\mu_{0,i} - 10)}.
\end{equation}

We denote Gaia EDR3 corrected parallax by $\varpi$, and perform a nonlinear least squares fit as
\begin{equation}
\varpi = 10^{-0.2(m_W + a_1(\log P - \log P_0) + a_2 + a_3\,\rm [Fe/H]) + 2} + zp_\varpi,
\end{equation}
where $zp_\varpi$ is the zero-point offset of the Gaia EDR3 parallax, and $a_1$, $a_2$, and $a_3$ are the coefficients for period, intercept, and metallicity, respectively. The parallax zero point is fixed to $zp_\varpi = 24 \pm 4\ \mu\mathrm{as}$, as discussed in Section~\ref{sec:floats}. We apply weighted least squares, using weights $1/e_{\rm para}^2$, where
\begin{equation}
e_{\rm para}^2 = \text{e\_plx}^2 + \varpi^2 \times \frac{0.08^2 + \sigma_m^2}{4 \times 1.086^2},
\end{equation}
with $\sigma_m$ denoting the uncertainty in apparent magnitude, and 0.08 representing the assumed intrinsic scatter of the PL relation. By fitting in parallax space, we determine the best-fit values of $a_1$, $a_2$, and $a_3$, as well as the dispersion $\sigma$ of the PLZ relation.

\begin{figure*}
\includegraphics[scale=0.65]{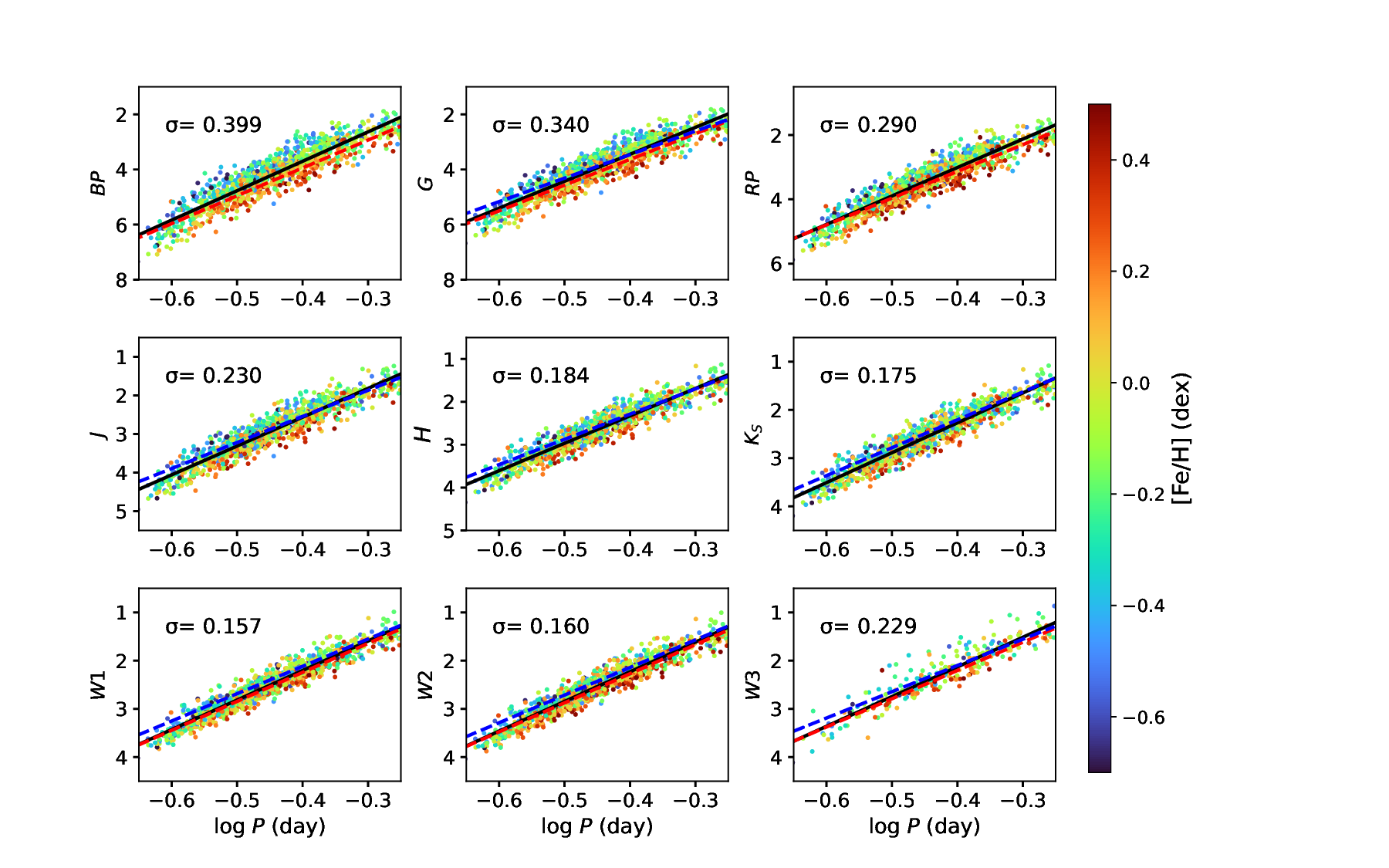}
\centering
\caption{Nine-bands PL relations of W UMa-type CBs. The black lines are the PL relations determined in the parallax space by our W UMa-type CBs. The red dashed lines are the PL relations from \protect\cite{song2024period} while the blue dashed lines are the PL relations from \protect\cite{chen2018optical}. The color bars represent metallicities.}
\label{fig1} 
\end{figure*}

\subsection{Absolute Magnitude}
In order to better present the derived the PL and PLZ relations in Figure \ref{fig1}, we convert the results obtained in parallax space to magnitude space by computing the absolute magnitude. To obtain more accurate absolute magnitudes, the effect of extinction on the absolute magnitudes in different bands should first be considered. Using the corrected parallax $\varpi$ and the parallax zero point \( zp_\varpi \), and we derived the absolute magnitude as
\begin{equation}
M_\lambda = m_\lambda - 5\log(\varpi-zp_\varpi) + 10 - A_\lambda \quad. 
\end{equation}
where $M_\lambda$, $m_\lambda$ and $A_\lambda$ are the apparent magnitude, apparent magnitude and extinction in each band. For W UMa-type CBs, the PL and PLZ relations are
\begin{equation}
M_\lambda = a_2 + a_1(\log P + 0.45), \quad 
\end{equation}
\begin{equation}
M_\lambda = a_2 + a_1(\log P + 0.45) + a_3[\text{Fe/H}]. \quad 
\end{equation}
$\log\,P_{0} = -0.45$ is the mean period. For early-type CBs, the PL and PLZ relations are
\begin{equation}
M_\lambda = a_2 + a_1(\log P + 0.10), \quad 
\end{equation}
\begin{equation}
M_\lambda = a_2 + a_1(\log P + 0.10) + a_3[\text{Fe/H}]. \quad 
\end{equation}
$\log\,P_{0} = -0.10$ is the mean period. \( a_1 \), \( a_2 \), and \( a_3 \) are the coefficients determined in the parallax space (see Section \ref{sec:2.2}).

\section{Results} \label{sec:style}
We present the results of PL and PLZ relations for W UMa-type and early-type CBs, obtained by using nonlinear least squares fitting based on samples with parallax uncertainties less than 2\%. Table \ref{tab1} lists the PL and PLZ relations for nine bands ($BP$, $G$, $RP$, $J$, $H$, $K_S$, $W1$, $W2$, $W3$) and five Wesenheit bands ($W_{H,K_S}$, $W_{J,K_S}$, $W_{J,H}$, $W_{W1,BP,RP}$, $W_{G,BP,RP}$) after fixing the parallax zero point \( zp_\varpi = 24 \pm 4 \) $\mu$as. \( a_1 \) represents the slope of the period, \( a_2 \) is the fitted intercept, \( a_3 \) indicates the coefficient of metallicity item.

\begin{figure*}
\includegraphics[scale=0.65]{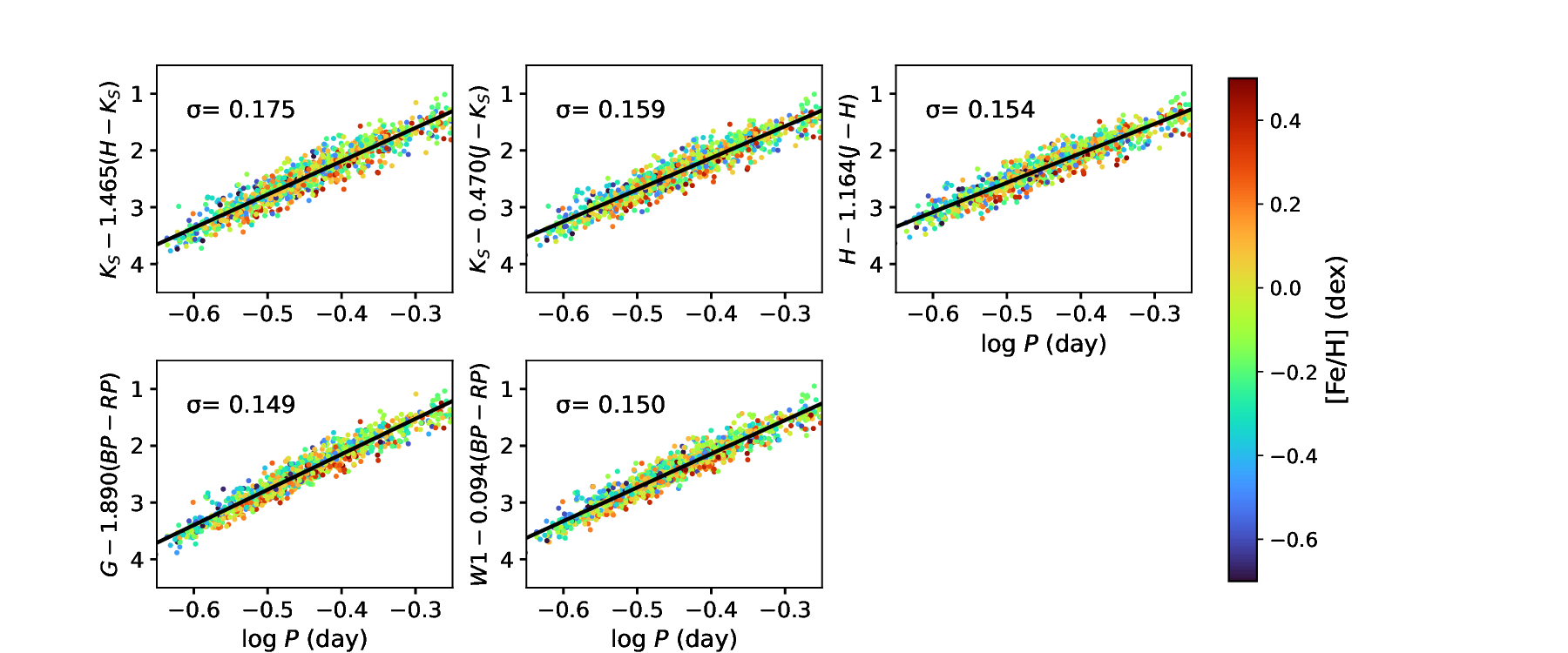}
\centering
\caption{Same as Figure \ref{fig1}, but this figure shows the Wesenheit PL relations of W UMa-type CBs.}
\label{fig2} 
\end{figure*}

\subsection{Period--Luminosity Relations} \label{sec:PLR}
The PL relation diagrams for W UMa-type CBs in these nine bands ($BP$, $G$, $RP$, $J$, $H$, $K_S$, $W1$, $W2$, $W3$) are shown in Figure \ref{fig1}, where \( \sigma \) represents the dispersion of the fitted PL relations. The black line in Figure \ref{fig1} represents our fit results, the red and blue dashed lines are from \cite{song2024period} and \cite{chen2018optical}, respectively. From the optical band to the infrared band, we compared the PL relation with the results of \cite{song2024period}. In the infrared band, our results are consistent with those of \cite{song2024period}, indicating that the PL relation of CBs is robust across different datasets. We also compared our findings with \cite{chen2018optical}, where the method for distinguishing W UMa-type CBs from early-type CBs is the same in both studies, based on a classification criterion of $\log P = -0.25$. Compared to their research, our constraints on dispersion are tighter, leading to a more precise PL relation. Figure \ref{fig2} shows the PL relations in the Wesenheit bands for W UMa-type CBs, where the Wesenheit bands can mitigate extinction effects and intrinsic scatter, resulting in more accurate PL relations. 

For W UMa-type CBs, the period slope \( a_1 \) increases continuously from optical bands to mid-infrared bands, while the intercept \( a_2 \) decreases with increasing wavelength. The dispersion of the PL relations decreases with increasing wavelength, with a minimum of 0.145 mag in the $W1$ band. This indicates that in single bands, the $W1$ band shows the best fit for the PL relations. For the five Wesenheit bands, \( W_{W1,BP,RP} \) and \( W_{G,BP,RP} \) have the smallest PL relations dispersion \( \sigma \) of 0.139 mag and 0.137 mag, respectively. The dispersion represents the error in the distance measured by one CB. For a spatially clumped object, such as a galaxy or cluster, which has multiple CBs, the distance error due to the dispersion of the PL relations gradually decreases to a value of $\sigma_{\rm PL}/\sqrt{n}$. At this point, systematic error terms become progressively more significant, such as the total zero point error of the PL relations \( zp_{\rm tot} \). It is the root-squared sum of the intercept error and the propagated error caused by parallax. We also list the total zero point error of the PL relations as one column in Table \ref{tab1} by Equation 3. For single band, the total error \( zp_{\rm tot} \) reaches its minimum value of 0.34\% in the $W1$ band. For the five Wesenheit bands, \( W_{W1,BP,RP} \) and \( W_{G,BP,RP} \) have the smallest total error of 0.33\%. This means that, ideally, the minimum error in distance calculated using these bands is about 0.3\%.

\begin{figure*}
\includegraphics[scale=0.65]{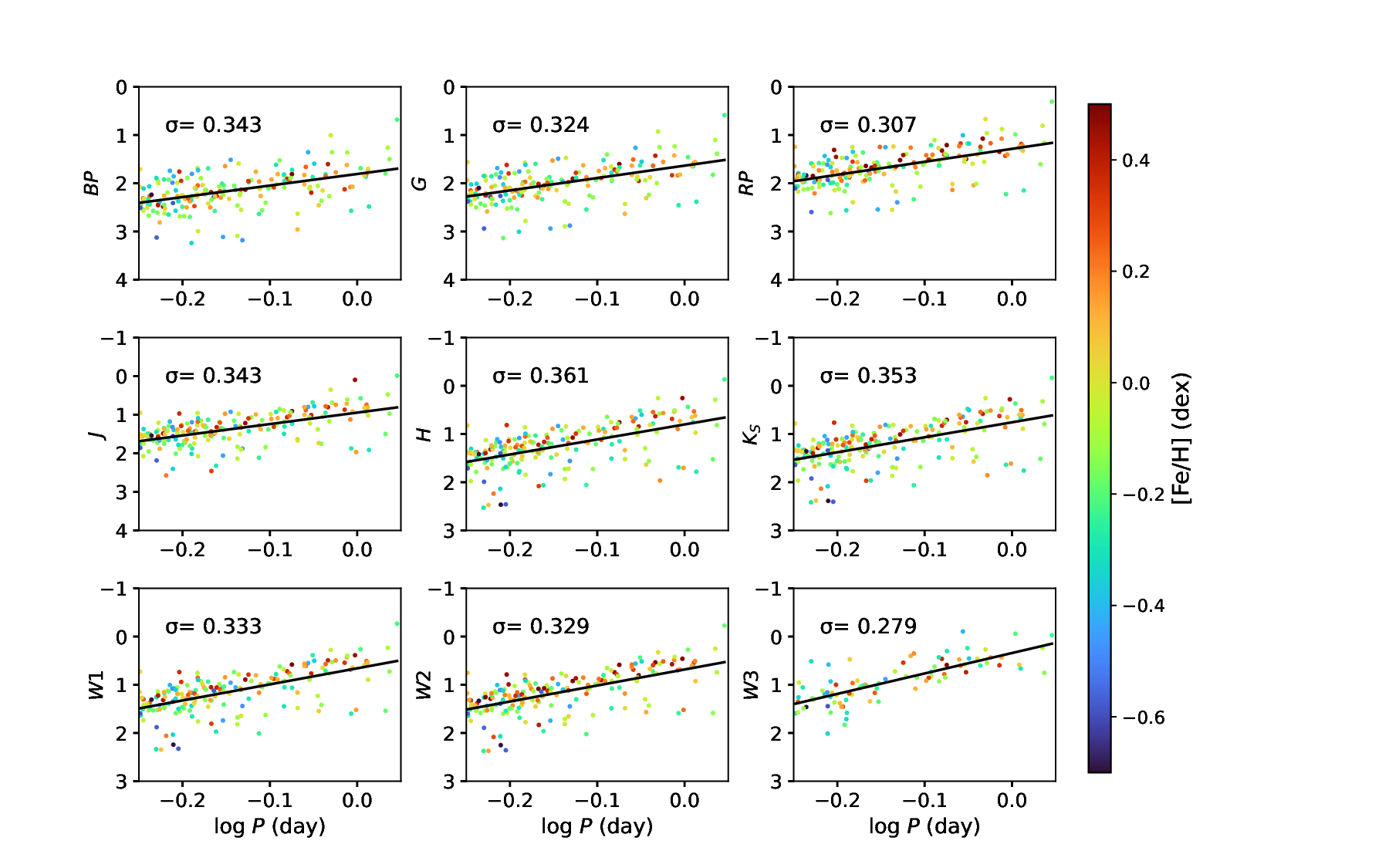}
\centering
\caption{The effect of metallicity on PL relations for nine bands of early-type CBs. The black lines show the PL relations we obtained for early-type CBs. The color bar represents the metallicity.}
\label{fig3} 
\end{figure*}

\begin{figure*}
\includegraphics[scale=0.65]{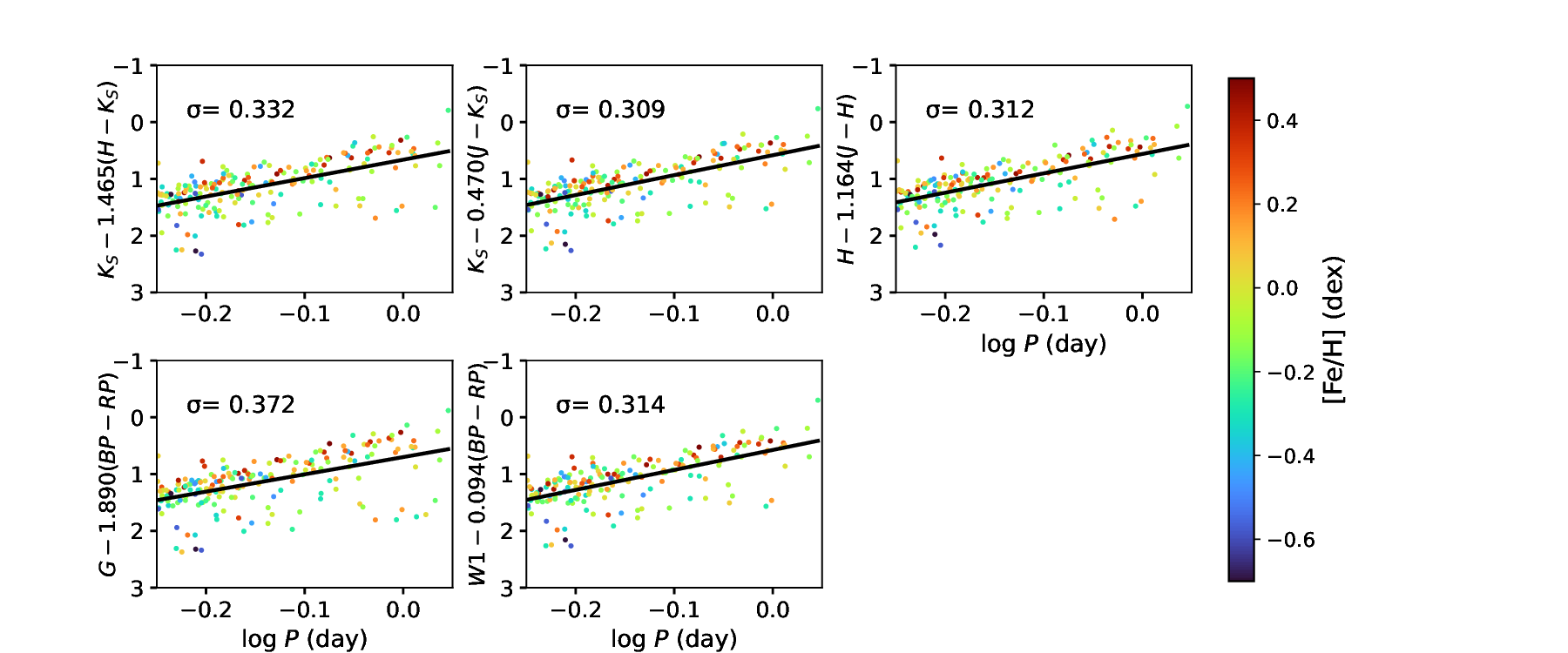}
\centering
\caption{Same as Figure \ref{fig3}, but this figure shows the Wesenheit PL relations of early-type CBs. }
\label{fig4} 
\end{figure*}

For early-type CBs, the period slope \( a_1 \) and intercept \( a_2 \) decrease continuously from optical bands to near-infrared bands. Figure \ref{fig3} shows the PL relations for early-type CBs in each single band. Figure \ref{fig4} shows the PL relations for early-type CBs in the Wesenheit bands. For a single band, the dispersion \( \sigma \) and total zero point error \( zp_{\rm tot} \) also decrease with increasing wavelength, reaching a minimum of 0.252 mag in the $W3$ band and 1.24\% in the $W2$ band. In $W1$, $W2$ and Wesenheit bands, the PL dispersion and \( zp_{\rm tot} \) are around 0.28 mag and 1.2\%, respectively, which are larger than those of W UMa-type CBs. This indicates that based on one early-type CB, the accuracy of the distance that can be obtained is 13\%.

\subsection{Period--Luminosity--Metallicity Relations} 
To investigate the effect of metallicity on PL relations in CBs, we derived the PLZ relations for each band using Equations 3 and 4, then transformed the results into magnitude space to show the PLZ relations in Figure \ref{fig1}. For the PLZ relations, the trends of the coefficients \( a_1 \), \( a_2 \) with wavelength are consistent with the PL relations.  The metallicity coefficient \( a_3 \) decreases from the optical bands to the infrared bands, reaching a minimum value of $0.154\pm0.021$ mag/dex. 
Regarding the impact of metallicity on the PL relation, our results indicate a weaker metallicity dependence compared to \cite{song2024period}, while the constraints on the dispersion remain consistent between the two studies.

The total zero point error \( zp_{\rm tot} \) and dispersion \( \sigma \) of PLZ relations also decrease with increasing wavelength, reaching a minimum of 0.34\% and 0.138 mag in the $W1$ band. When considering the effect of metallicity on the PL relations, the dispersion \( \sigma \) has improved significantly. When the metallicity [Fe/H] increases, the absolute magnitude becomes faint. In the Wesenheit bands, \( W_{W1,BP,RP} \) and \( W_{G,BP,RP} \) exhibit the minimum \( zp_{\rm tot} \) of 0.33\%, and minimum dispersion \( \sigma \) of 0.134 mag and 0.133 mag, respectively. Therefore, with a parallax uncertainty of 2\%, we found three bands with the best PLZ relations, namely $W1$, \( W_{W1,BP,RP} \), and \( W_{G,BP,RP} \).

For early-type CBs, the period slope \( a_1 \) and intercept \( a_2 \) decrease continuously from optical bands to mid-infrared bands, while the metallicity slope remains relatively stable at approximately $-0.203$ mag/dex. With increasing wavelength, \( zp_{\rm tot} \) and \( \sigma \) reach the best values of 1.23\% in the $W2$ band and 0.252 mag in the $W3$ band, respectively. In terms of the Wesenheit bands, \( zp_{\rm tot} \) and \( \sigma \) are similar to the infrared bands.

\begin{figure*}
\includegraphics[scale=0.55]{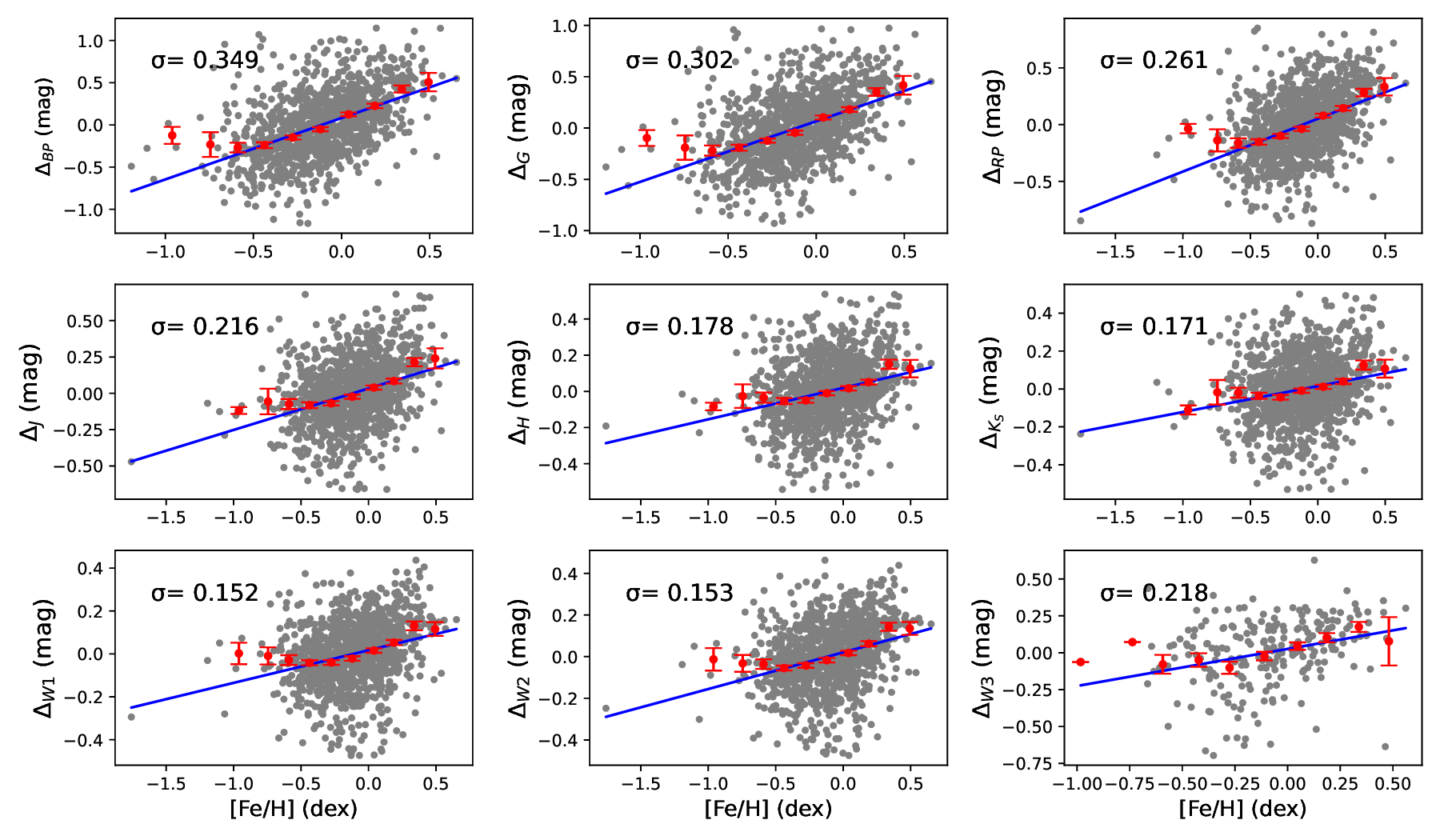}
\centering
\caption{Best-fit relations for metallicity and absolute magnitude residuals for W UMa-type CBs. The ten red dots are the mean values in different metallicity ranges as well as their error bars. }
\label{fig5} 
\end{figure*}

\begin{figure*}
\includegraphics[scale=0.55]{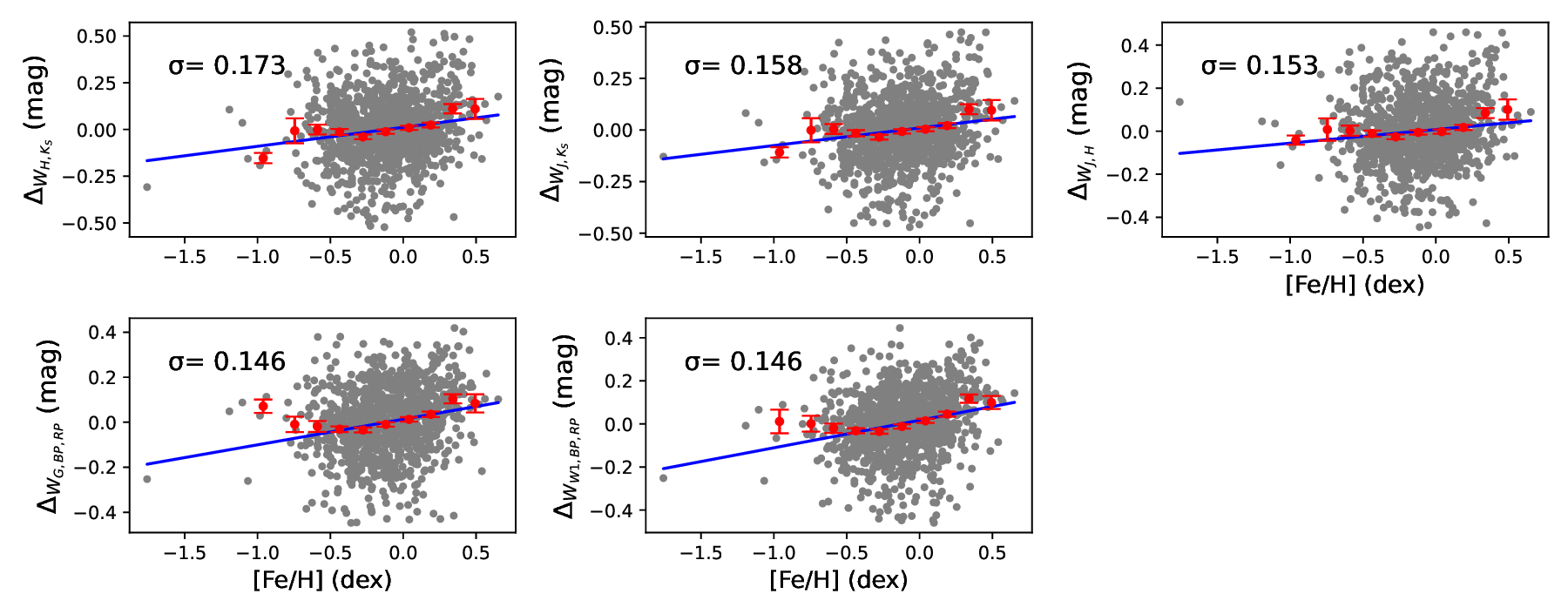}
\centering
\caption{Best-fit relations for metallicity and absolute magnitude residuals for the Wesenheit bands for W UMa-type CBs. The ten red dots are the mean values in different metallicity ranges as well as their error bars.}
\label{fig6} 
\end{figure*} 

Based on the PL relations obtained in Section \ref{sec:PLR}, we can calculate the absolute magnitude residuals for each band. Then, we can analyze the relationship between the metallicities and the absolute magnitude residuals. This gives a better demonstration of how the metallicity affects the PL relations in each interval compared to obtaining the PLZ relations directly. We represent the absolute magnitude residuals as \( \Delta \), where \( \Delta_{\lambda} = M - M_{\lambda, PL} \), with \( M \) being the observed absolute magnitude and \( M_{\lambda, PL} \) being the absolute magnitude determined from PL relations. Figure \ref{fig5} displays the distribution of the residuals of absolute magnitude $\Delta_{\lambda}$ with metallicity [Fe/H] for W UMa-type CBs. The absolute magnitude residuals increase with increasing metallicity. From optical bands to mid-infrared bands, metallicity has less and less effect on PL relations. In the Wesenheit bands, the metallicity effect is even smaller than in the mid-infrared band, as shown in Figure \ref{fig6}. This indicates that if the metallicity is not known, it is more accurate to use the PL relations in the infrared bands or the Wesenheit bands to calculate distances of CBs.

\begin{figure*}
\includegraphics[scale=0.55]{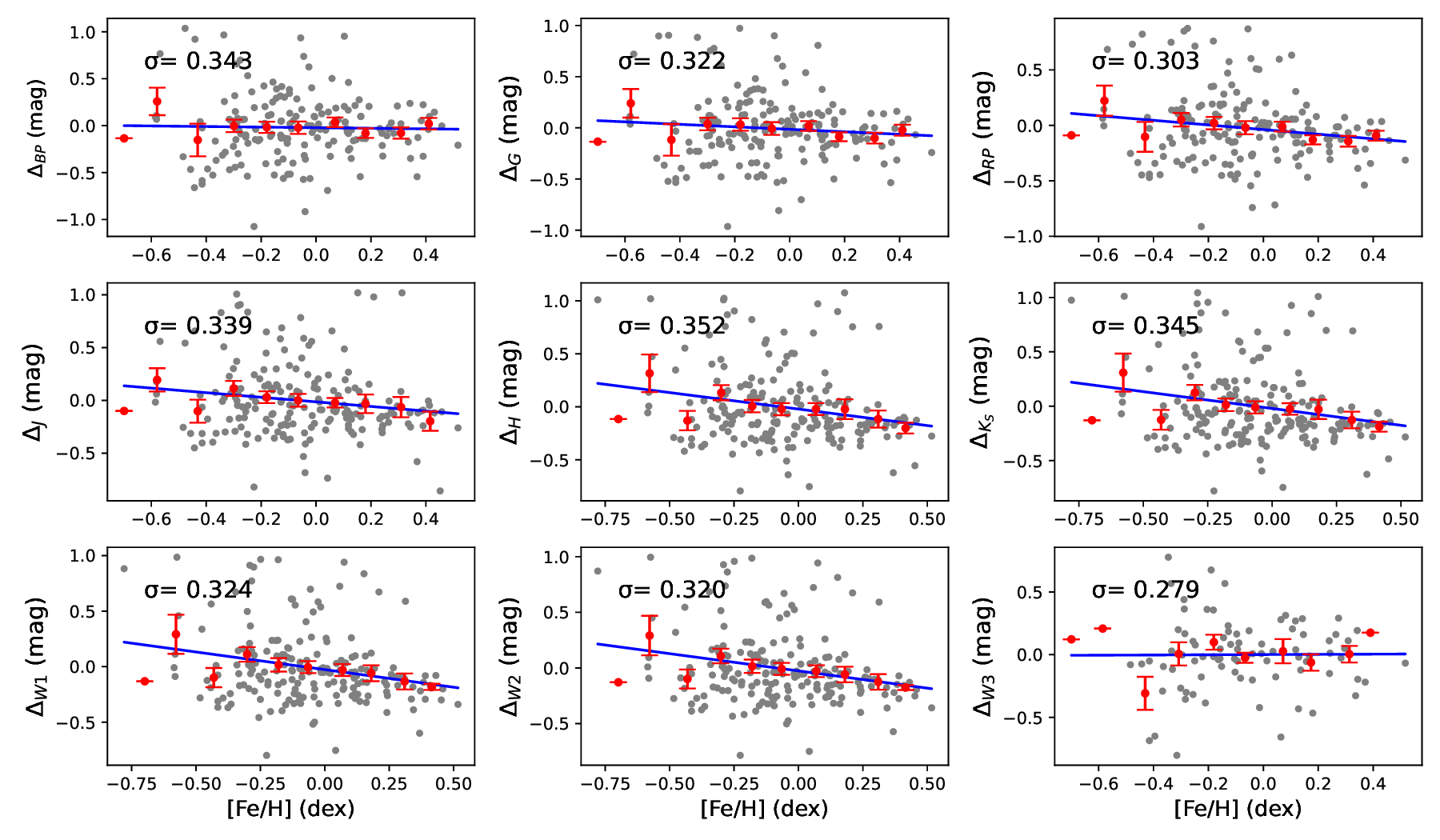}
\centering
\caption{Best-fit relations for metallicity and absolute magnitude residuals for early-type CBs. The ten red dots are the mean values in different metallicity ranges as well as their error bars.}
\label{fig7} 
\end{figure*}

For early-type CBs, there is no clear trend in the absolute magnitude residuals versus metallicity distribution from Fig. \ref{fig7} due to the large dispersion. Figure \ref{fig8} shows the metallicity versus absolute magnitude residuals in the Wesenheit band at $\log P>-0.25$. The effect of metallicity on the early-type CB PL relations is uncertain.

\section{Discussions} 
As distance indicators, the accuracy of distance measurements using CBs can be assessed through the dispersion of the PL relation, the zero-point uncertainty of the PL relation, and the metallicity dependence of the PL relation (also known as the metallicity effect). The dispersion of the PL relation reflects the distance accuracy for an individual CB. The zero-point uncertainty and the metallicity effect, on the other hand, determine the lower limit of the distance error when using a large number of CBs to measure the distance to an object such as a star cluster or a galaxy. Only through careful analysis of Gaia parallax uncertainties can the statistical and systematic errors of the CB PL relation be optimized. In this section, we first discuss how we handle the analysis of Gaia parallax uncertainties, followed by a discussion of the statistical and systematic errors of the CB PL relation.

\label{sec:floats}
\subsection{Gaia parallax zero point}
Gaia's parallax was found to have a possible offset that affects the PL relations of the CB. The use of W UMa-type CBs with different parallaxes breaks the degeneracy of the PL relation zero point \( zp_{\rm tot} \) and the parallax zero point \( zp_\varpi \). To determine the parallax zero point, we consider samples fulfill different parallax uncertainty criteria. Using Equations 3 and 4, we can derive the intercept \( a_2 \), the  metallicity effect \( a_3 \), and the parallax zero point \( zp_\varpi \) for samples under different parallax uncertainty criteria. We choose the three best PLZ relations, namely $W1$, \( W_{W1,BP,RP} \), and \( W_{G,BP,RP} \), to constrain the intercept and metallicity effect and the parallax zero point \( zp_\varpi \).

 \begin{figure*}
\includegraphics[scale=0.75]{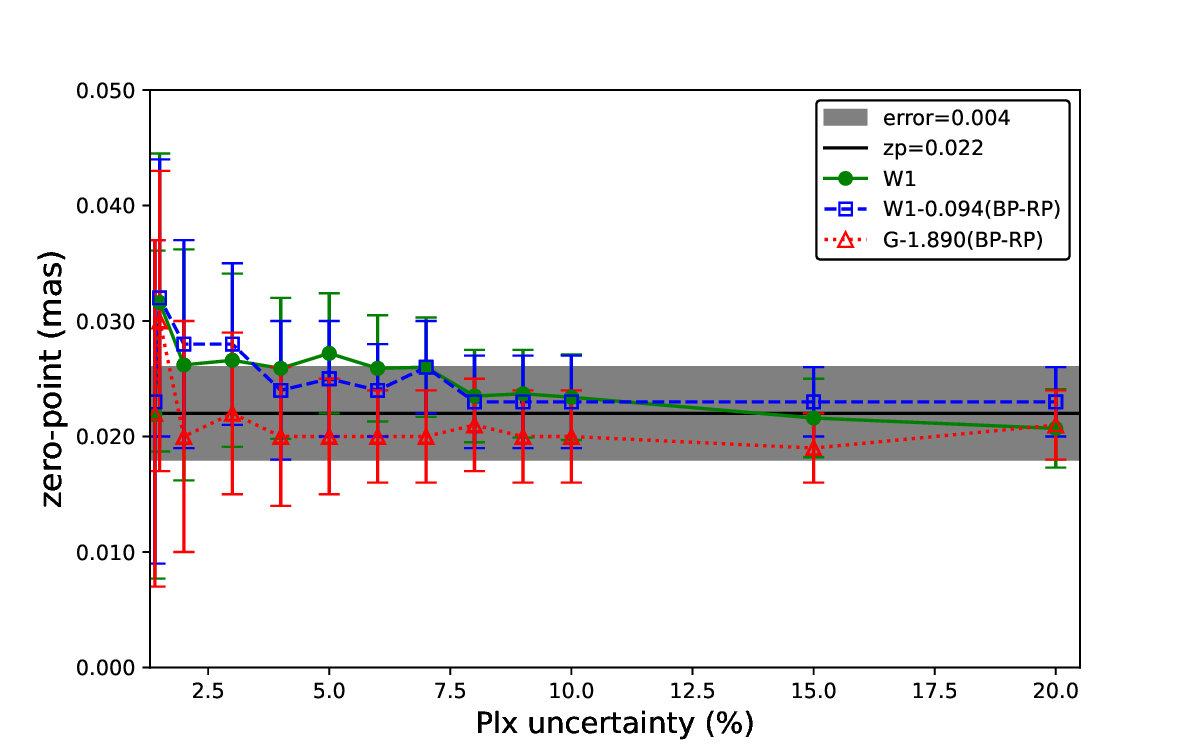}
\centering
\caption{The parallax zero point values \( zp_\varpi \) determined from three bands under different parallax uncertainties.}
\label{fig9} 
\end{figure*}

Figure \ref{fig9} illustrates the variation of the parallax zero point \( zp_\varpi \) with the parallax uncertainty threshold. We find that the parallax zero points \( zp_\varpi \) of all three PL relations are stable and have small errors when the parallax uncertainty threshold changes from 20\% to 8\%. We take the average of the determined parallax zero point \( zp_\varpi \) within this interval and obtain \( zp_\varpi = 24 \pm 4 \) $\mu$as. As the parallax uncertainty threshold falls below 8\% and continues to decrease, the error bar in \( zp_\varpi \) increases, indicating that the sample size gradually decreases and is unable to constrain \( zp_\varpi \) effectively. The shaded area in Figure \ref{fig9} represents the $\pm 1 \sigma$ distribution around the average parallax zero point \( zp_\varpi \). We found the shaded region to be suitable for samples with any parallax uncertainty threshold. Therefore, in this study, we use \( zp_\varpi = 24 \pm 4 \) $\mu$as to determine the PL and PLZ relations of CBs. We find that the CB-based parallax zero point is in agreement with that determined by Cepheids \( zp_\varpi = 14 \pm 6 \) $\mu$as in the 2$\sigma$ \citep{riess2021cosmic}.

\subsection{Accuracy of W UMa-type CB Luminosity}

As standard candles, we need to evaluate the accuracy of the W UMa-type CB PL and PLZ relations, where the two most important quantities are dispersion and total zero point error \( zp_{\rm tot} \). The dispersion represents the error in the distance measured by one W UMa-type CB, while the total zero error represents the error in the ideal distance that can be obtained based on an infinite number of W UMa-type CBs. We derived the total zero error as
\begin{equation}
zp_{\rm tot} \ = \sqrt{\left(\frac{a_{2,std}}{2/1.086}\right)^2 + \left(\frac{2}{\text{mean}(\bar{\omega})/1000}\right)^2} \quad. 
\end{equation}
where \( a_{2,std} \) denotes the standard
deviation of the intercept term in the PLR fit, and the second term represents the contribution from parallax
uncertainty propagation.

From Table \ref{tab1}, the PLZ relations dispersion decreases from optical band to the infrared band, and achieve 6\%-7\% in mid-infrared bands and Wesenheit bands. Note that we use the sample with a parallax uncertainty of less than 2\%, and the dispersion of the PL relations still decreases when this threshold is reduced. \cite{song2024period} also determined the W UMa-type CB PLZ relations with the dispersion of 6\%.

\begin{figure*}
\includegraphics[scale=0.75]{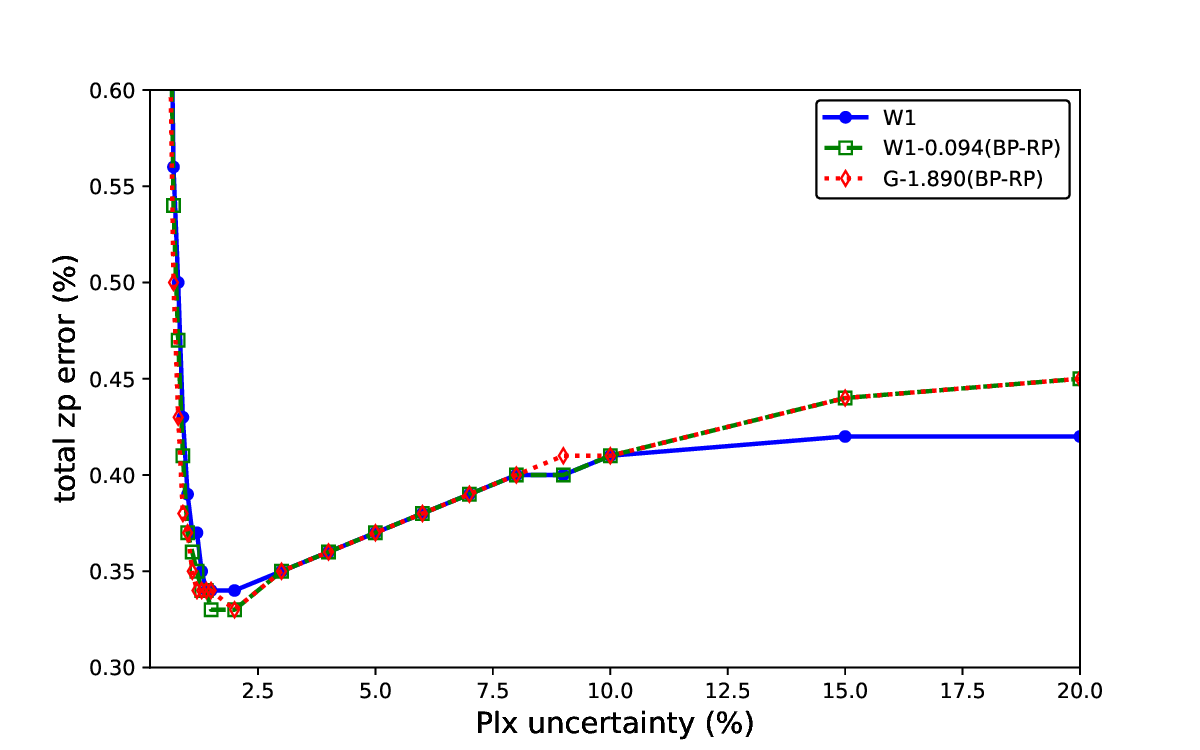}
\centering
\caption{The total zero point error \( zp_{\rm tot} \) for $W1$ and two Wesenheit bands under different parallax uncertainties. The blue line represents the $W1$ band, the green dotted line represents the $W_{W1,BP,RP}$ band, and the red dotted line represents the $W_{G,BP,RP}$ band.}
\label{fig10} 
\end{figure*}

When we use multiple W UMa-type CBs to measure a cluster or galaxy distance, we are particularly concerned with the total zero point error, {\bf which has not been addressed in previous studies in terms of optimization.} The total zero point error \( zp_{\rm tot} \) includes both the PLZ relations intercept error and the propagated error from the zero point error of Gaia parallax. As the parallax uncertainty threshold increases, the number of W UMa-type CBs used for the analysis increases, and the mean distance increases. At this point the intercept error decreases slightly, while the propagated error increases significantly. We also select the $W1$, $W_{W1,BP,RP}$, and $W_{G,BP,RP}$ bands, which exhibit the best PLZ relations, to determine the total zero point error of the PLZ relation. Figure \ref{fig10} shows how \( zp_{\rm tot} \) vary with the parallax uncertainty threshold, where the blue line represents the $W1$ band, the green and red dotted lines represent the $W_{W1,BP,RP}$ and $W_{G,BP,RP}$ bands. As shown in Figure \ref{fig10}, the total zero point errors in the PLZ relation first decrease and then increase with increasing parallax uncertainty threshold. The total PLZ relations zero point errors are minimized for a parallax uncertainty threshold of 2\%, with 0.34\%, 0.33\%, and 0.33\% for the $W1$, $W_{W1,BP,RP}$, and $W_{G,BP,RP}$ bands, respectively. \( zp_{\rm tot} \) of CB PLZ relations is much smaller than Cepheids \citep[0.7\%,][]{2022ApJ...934L...7R} due to a larger nearby sample ($\sim $1000 W UMa-type CBs with the mean distance of 702 pc). The $\sim 0.3$\% value implies that if an object contains an infinite number of CBs, the distance uncertainty to that object can be optimized to $\sim 0.3$\%. This suggests that W UMa-type CBs are very promising distance tracers that could be used in the future to more accurately measure the distances to dwarf galaxies.

\subsection{PL relations in different metallicity intervals}
In this section, we discuss in detail the effect of metallicity on the absolute magnitude residuals between the different metallicity intervals. We fitted the PL relations for nine individual bands and five Wesenheit bands for W UMa-type CBs. Figures \ref{fig5} and \ref{fig6} show the relations between metallicity and absolute magnitude residuals for W UMa-type CBs. We divided metallicity into ten intervals, each with a width of 0.16 dex. The red dots represent the average metallicity and the average absolute magnitude residuals within each interval. The error bars are the standard deviation within each interval. From Figures \ref{fig5} and \ref{fig6}, it can be seen that the residuals within each interval gradually increases with increasing metallicity, which is consistent with the results we derived in Section \ref{sec:style}.
However, the results in some metallicity intervals slightly deviate from the overall fitting results. We chose the $W1$ band, which showed the best performance in the single-band analysis, to calculate the deviations in different metallicity intervals. When the metallicity is $\mathrm{[Fe/H]} > -0.68$ dex, the maximum deviation between the red dots and the fitted line is 0.062 mag. These deviations are within 3$\sigma$ errors. When the metallicity is poor with $\mathrm{[Fe/H]} < -0.68$ dex, the deviations between the red dots and the fitted line are 0.131 mag and 0.087 mag. For W UMa-type CBs that are too metal-poor, the deviation from  PLZ relations can be explained by errors ($\sim$ 0.05 mag), i.e., the number of W UMa-type CBs with extreme metallicity is too small. This suggests that it is suitable to use a constant metallicity coefficient for W UMa-type CBs.

 \begin{figure*}
\includegraphics[scale=0.55]{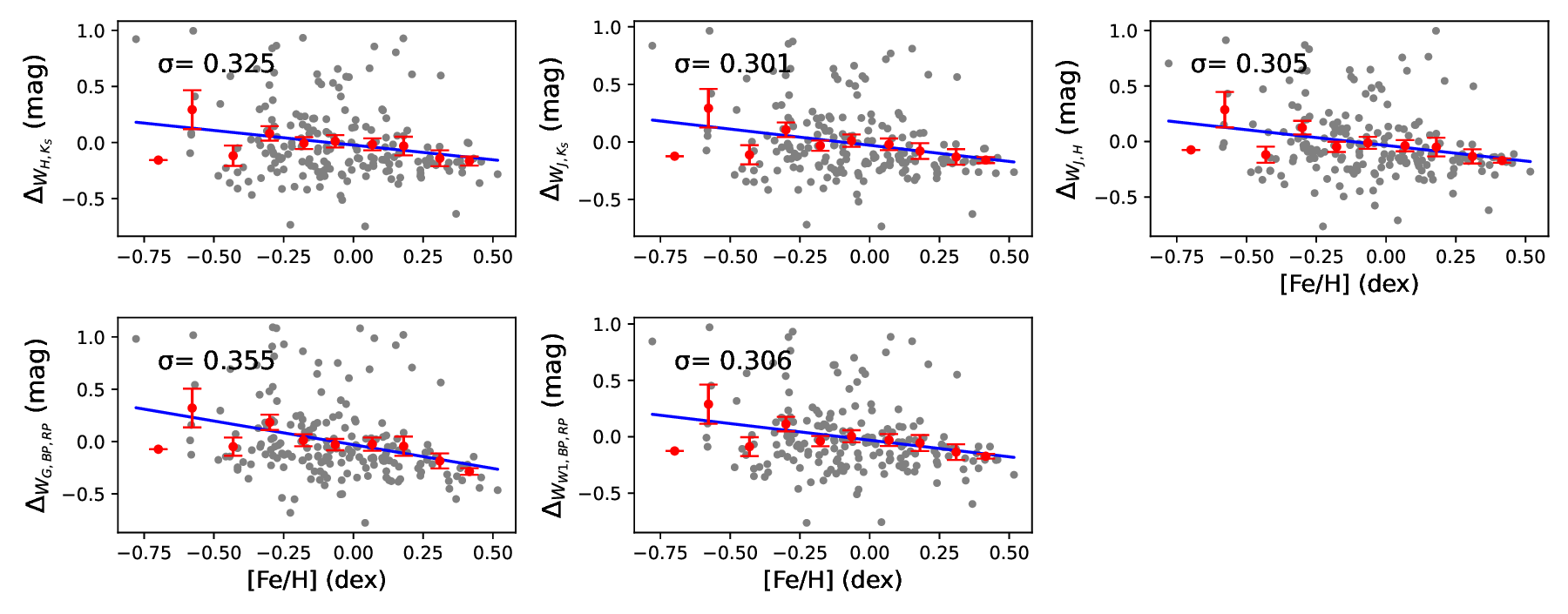}
\centering
\caption{Best-fit relations for metallicity and absolute magnitude residuals for the Wesenheit bands for early-type CBs. The ten red dots are the mean values in different metallicity ranges as well as their error bars. }
\label{fig8} 
\end{figure*}   

Figures \ref{fig7} and \ref{fig8} illustrate the fitted plots for metallicity and absolute magnitude residuals for early-type CBs. Overall, the residuals are weakly correlated with metallicity. When the metallicity is within $-0.375 < \mathrm{[Fe/H]} < 0.500$ dex, the maximum deviation between the red dots and the fitted line is 0.041 mag. When the metallicity $\mathrm{[Fe/H]} < -0.375$ dex, the deviations between the red dots and the fitted line are around 0.2 mag. However, these deviations can also be explained by errors. 

\section{Conclusions} 
\label{sec:5}
Using Gaia EDR3 parallax, LAMOST metallicity, we establish PL and PLZ relations for W UMa-type CBs and early-type CBs in nine single bands (from optical to the mid-infrared) and five Wesenheit bands. In order to avoid non-normal distribution of parallax errors, we performed determination in parallax space. Using the three best PLZ relations, we determined that for CB, the Gaia-corrected parallax zero offset is $zp_\varpi=24\pm4$ $\mu$as. Using different Gaia parallax selection criteria, we found that the CB PL and PLZ relations with the smallest total zero point error can be obtained when using a parallax percentage error of less than 2\%.

For W UMa-type and early-type CBs, the dispersion and total zero error of the PL relation and PLZ relation decrease from the optical band to the mid-infrared band. Therefore, it is most appropriate to use either the infrared band or the Wesenheit band photometry to measure the CB distance. For W UMa-type CBs, the metallicity effect is significant and it is better to use the PLZ relation when metallicity information is available. Based on a single CB, the distance error can be as low as 6\% (0.13 mag) and 12\% (0.27 mag) for W UMa-type and early-type CB, respectively.

W UMa-type CBs are very good distance tracers because of the higher number of and smaller total zero point error. When there are enough CBs, the distance errors obtained for galaxies or clusters can be as low as 0.33\%, which is less than half the distance error based on classical Cepheids. We also checked whether the PLZ relation is consistent across different metallicity intervals and found that only in very metal-poor intervals there is some deviation from the overall PLZ relation, but that deviation can be explained by the errors. In the future, based on observations from telescopes such as the Vera C. Rubin Observatory Legacy Survey of Space and Time (LSST) and China Space Station Telescope (CSST), W UMa-type CBs will be used to measure very high-precision distances to galaxies in the Local Group such as M31 and M33.

\section*{Acknowledgements}
We thank the anonymous referee for the helpful comments. This work was supported by the National Natural Science Foundation of China (NSFC) through grants 12173047, 12373035, 12322306, 12373028, 12233009, 12273018. We also thanked the support from the National Key Research and development Program of China, grants 2022YFF0503404. X. Chen and S. Wang acknowledge support from the Youth Innovation Promotion Association of the Chinese Academy of Sciences (no. 2022055 and 2023065). This publication is based on observations obtained with the Guoshoujing Telescope (the Large Sky Area Multi-Object Fiber Spectroscopic Telescope; LAMOST). LAMOST is a National Major Scientific Project constructed by the Chinese Academy of Sciences. Funding for the project has been provided by the Chinese National Development and Reform Commission. LAMOST is operated and managed by the National Astronomical Observatories, Chinese Academy of Sciences. This work presents results from the European Space Agency (ESA) space mission Gaia. Gaia data are being processed by the Gaia Data Processing and Analysis Consortium (DPAC). Funding for the DPAC is provided by national institutions, in particular the institutions participating in the Gaia MultiLateral Agreement (MLA). The Gaia mission website is \url{https://www.cosmos.esa.int/gaia}. The Gaia archive website is \url{https://archives.esac.esa.int/gaia}.

\section*{DATA AVAILABILITY}
The data underlying this article will be shared on reasonable request to the corresponding author.

\bibliographystyle{mnras}
\bibliography{ref} 

\bsp	
\label{lastpage}
\end{document}